\documentclass[preprint,12pt]{elsarticle}

\usepackage{graphicx}
\usepackage{amssymb}
\journal{\ }

\begin{document}

\begin{frontmatter}


% use the tnoteref command within \title for footnotes;
% use the tnotetext command for theassociated footnote;
% use the fnref command within \author or \address for footnotes;
% use the fntext command for theassociated footnote;
% use the corref command within \author for corresponding author footnotes;
% use the cortext command for theassociated footnote;
% \author{Name\corref{cor1}\fnref{label2}}
% \address{Address\fnref{label3}}

\title{Study of Beam Profile Measurement at Interaction Point in
 International Linear Collider}


\author[label1]{Kazutoshi Ito}
 \ead{kazuto@awa.tohoku.ac.jp}
 \author[label2]{Akiya Miyamoto}
 \author[label1]{Tadashi Nagamine}
 \author[label2]{Toshiaki Tauchi}
 \author[label1]{Hitoshi Yamamoto}
 \author[label1]{Yosuke Takubo}
 \author[label1]{Yutaro Sato}

\address[label1]{Department of Physics, Tohoku University, Sendai,
 Japan}
\address[label2]{High Energy Accelerator Research Organization, KEK, 1-1
 Oho, Tsukuba, Ibaraki 305-0801, Japan}

\begin{abstract}
% Text of abstract
 At the international linear collider,
 measurement of the beam profile at the interaction point is a key issue
 to achieve high luminosity.
 We report a simulation study on a new beam profile monitor, called the
 pair monitor, which uses the hit distribution of the electron-positron
 pairs generated at the interaction point.
 We obtained measurement accuracies of 5.1\%, 10.0\%, and
 4.0\% for the horizontal ($\sigma_x$),
 vertical ($\sigma_y$), and 
 longitudinal beam size ($\sigma_z$), respectively, for 50 bunch
 crossings.
\end{abstract}

\begin{keyword}
% MSC codes here, in the form: \MSC code \sep code
% or \MSC[2008] code \sep code (2000 is the default)
ILC, beam profile, pair monitor, interaction point
\end{keyword}

\end{frontmatter}

% main text
\section{Introduction}
\label{introduction}
The International Linear Collider (ILC) is the next-generation
electron-positron collider  at the high energy frontier.
The total length of the main linac is about 31 km. The center of
mass energy is 500 GeV at the first stage.
The beam bunch consists of $2.05\times10^{10}$ particles,
and its size at the interaction point (IP) is 639 nm (width) $\times$
5.7 nm (height) $\times$ 300 $\mathrm{\mu m}$ (length) to achieve a luminosity
of $2\times10^{34}\ \mathrm{cm}^{-2}\mathrm{s}^{-1}$.
A beam train consists of 2625 bunches, and the train is
repeated at 5 Hz.
The nominal beam parameters for ILC are given in Table 
\ref{tbl:nominalBeamParameter} \cite{RDR}.

At ILC, measurement of the beam size at IP is essential
since the luminosity critically depends on beam size as:
\begin{eqnarray}
 {\cal L} = \frac{1}{4\pi}\frac{f_{rep}n_bN^2}{\sigma_x\sigma_y} \times
  H_D, 
  \label{luminosity}
\end{eqnarray}
where $f_{rep}$ is the train repetition rate per second, $n_b$ is the number of beam bunches
per train,
$N$ is the number of the particles per beam bunch, $\sigma_x\ (\sigma_y)$
is the horizontal (vertical) beam size and
$H_D$ is the disruption enhancement factor (typically $H_D \sim 2$)
\cite{phd}.
The vertical beam size is very small,
and it must be measured with about 1 nm accuracy
\cite{TauchiAndYokoya}. In addition, the space to locate the beam
profile monitor is limited. To satisfy those requirements,
we study a new beam profile monitor called the pair monitor which
utilizes the large number of electron-positron pairs created at
IP.

With the beam energy and the particle density of ILC, 
a large number of $e^+e^-$ pairs
are created during the bunch crossing by the following three
incoherent processes; Breit-Wheeler process ($\gamma + \gamma
\rightarrow e^- + e^+ $), Bethe-Heitler process ($\gamma + e
\rightarrow e + e^- + e^+$) and Landau-Lifshitz process ($e + e
\rightarrow e + e + e^- + e^+$), where $\gamma$ is a beam-strahlung
photon \cite{phd}. The generated $e^\pm$ pairs are usually 
referred to as the pair background.
The particles with the same charge as the oncoming beam
are scattered with large angles and carry information on the beam
profile \cite{TauchiAndYokoya, pairCreation}.

The pair monitor measures the beam profile by using the azimuthal
distributions of the scattered $e^+e^-$ pairs \cite{takubo}.
In this paper, we report a reconstruction of the beam sizes using the
Taylor matrixes and present the expected measurement accuracies.

\section{Simulation}
\label{setup}
\begin{figure}[htbp]\centering
 \includegraphics[width=11cm]{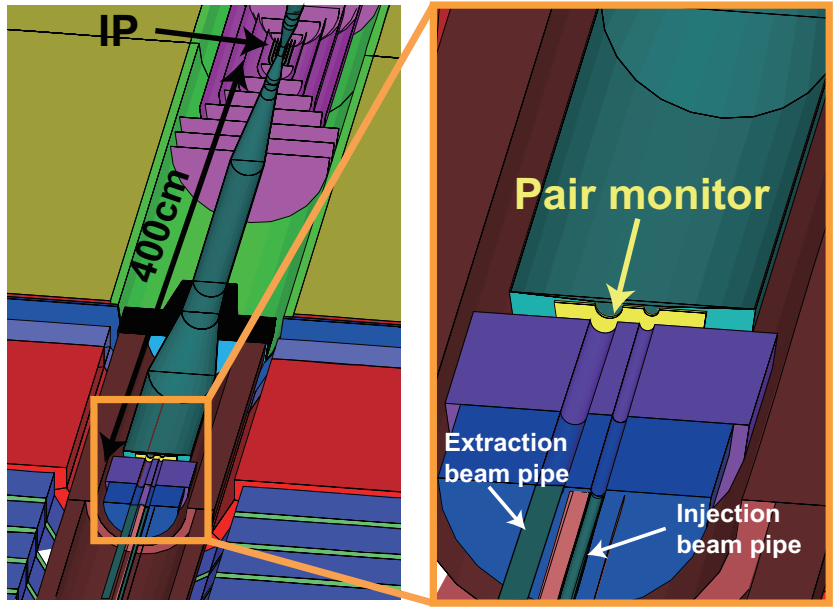}
 \caption{The detector geometry and location of the pair monitor. The
 pair monitor is located at 400 cm from IP.}
 \label{fig:location}
\end{figure}

The performance of the pair monitor was studied using the geometry of
the GLD detector \cite{Jupiter}. The pair background was
generated by CAIN \cite{CAIN} assuming head-on beam bunch collision..
The pair monitor was located at 400 cm from IP as shown in Figure
 \ref{fig:location}.
Solenoid field (3T) with the anti-DID (reversed Detector Integrated Dipole)
\cite{RDR} was used for the magnetic field.
The anti-DID is a correction coil wound on the main
solenoid. It is designed to lead the
pair backgrounds to the extraction beam pipes so that detector
backgrounds can be minimized.
The pair monitor is a silicon disk of 10 cm
radius and 200 $\mathrm{\mu m}$ thickness. There are two holes whose radius are 1.0 cm and 1.8 cm for
the incoming and outgoing beams, respectively.

\begin{table}[htbp]\centering
 \begin{tabular}{|c||c|c|}
  \hline
  Parameter & Unit &  \\ \hline \hline
  Center of mass energy & GeV & 500 \\ \hline
  Number of particles per bunch & $\times 10^{10}$ & 2.05\\ \hline
  Number of bunches per train & & 2625 \\ \hline
  Train repetition & Hz & 5\\ \hline
  Normalized horizontal emittance at IP & mm-mrad & 10\\ \hline
  Normalized vertical emittance at IP & mm-mrad & 0.04\\ \hline
  Horizontal beta function at IP & mm & 20\\ \hline
  Vertical beta function at IP & mm & 0.4\\ \hline
  Horizontal beam size at IP & nm& 639\\ \hline
  Vertical beam size at IP & nm& 5.7\\ \hline
  Longitudinal beam size at IP & $\mu m$& 300\\ \hline
  Crossing angle & mrad& 14\\ \hline
 \end{tabular}
 \caption{The nominal beam parameters for ILC.}
\label{tbl:nominalBeamParameter}
\end{table}

\section{The reconstruction method of the beam size}
\label{reconstruction}
We reconstructed the beam sizes from the hit distribution of the pair
backgrounds at the pair monitor. The measurement variables used for the
reconstruction were, the shoulder radius of the hit distribution, the
number of hits in two regions of pair monitor, and the total number of
the hits.
Since these measurement variables ($m_i$, $i=1,
2, \cdots, n$) should depend
on the beam sizes ($\sigma_x, \sigma_y, \sigma_z$), they can be expanded around the nominal beam sizes
($\sigma_x^0, \sigma_y^0, \sigma_z^0$) by the Taylor expansion as
follows.
\begin{eqnarray}
 \label{Delta_Mi}
\Delta m_i &=& 
  m_i(\sigma_x, \sigma_y, \sigma_z) - m_i(\sigma_x^0, \sigma_y^0,
  \sigma_z^0)
  \nonumber\\
 &=&\sum_{\alpha=x,y,z}\frac{\partial m_i}{\partial \sigma_\alpha}\Delta\sigma_\alpha
 +
 \sum_{\alpha=x,y,z} \sum_{\beta=x,y,z}\frac{1}{2}\Delta\sigma_\beta\frac{\partial^2m_i}{\partial\sigma_\alpha\partial\sigma_\beta}\Delta\sigma_\alpha
 + \cdots \nonumber \\
 &=&\sum_{\alpha=x,y,z}\left [\frac{\partial
     m_i}{\partial\sigma_\alpha}+\frac{1}{2} \sum_{\beta=x,y,z}\Delta\sigma_\beta\frac{\partial^2m_i}{\partial\sigma_\alpha\partial\sigma_\beta}+\cdots\right]\cdot\Delta\sigma_\alpha, 
\end{eqnarray}
where $\Delta m_i = m_i(\sigma_x, \sigma_y, \sigma_z) - m_i(\sigma_x^0,
\sigma_y^0, \sigma_z^0), \Delta\sigma_\alpha =
\sigma_\alpha-\sigma_\alpha^0$.
 Equation (\ref{Delta_Mi}) can be expressed by using
vectors and matrixes as
\begin{eqnarray}
\label{matrix}
\Delta\vec{m}=\left[A_1+\Delta\vec{\sigma}^T\cdot
	       A_2+\cdots\right]\cdot\Delta\vec{\sigma},
\end{eqnarray}
where $\Delta \vec{m}=(\Delta m_1, \Delta m_2, \cdots, \Delta m_n),
\Delta\vec{\sigma} =
(\Delta\sigma_x, \Delta\sigma_y, \Delta\sigma_z)$ and $A_1$ is a
$n\times3$ matrix of the first order coefficients of the Taylor expansion and $A_2$
is a tensor of the second derivative coefficients. The beam size is reconstructed by
multiplying the inverted matrix of a coefficient of $\Delta\vec{\sigma}$ in
Equation (\ref{matrix}) as follows.
\begin{eqnarray}
 \label{inverseMatrix}
 \Delta\vec{\sigma}=\left[A_1+\Delta\vec{\sigma}^T\cdot
		     A_2+\cdots\right]^+\cdot\Delta\vec{m},
\end{eqnarray}
where the superscript ``+'' indicates the Moore Penrose inversion which
gives the inverse matrix of a non-square matrix $A$ as $A^+ =
(A^TA)^{-1}A^T$ \cite{diagnostics, fastBeamDiagnostics}.

\section{The measurement variables}
\label{measurementVariables}
\begin{figure}[htbp]\centering
 \includegraphics[width=11cm]{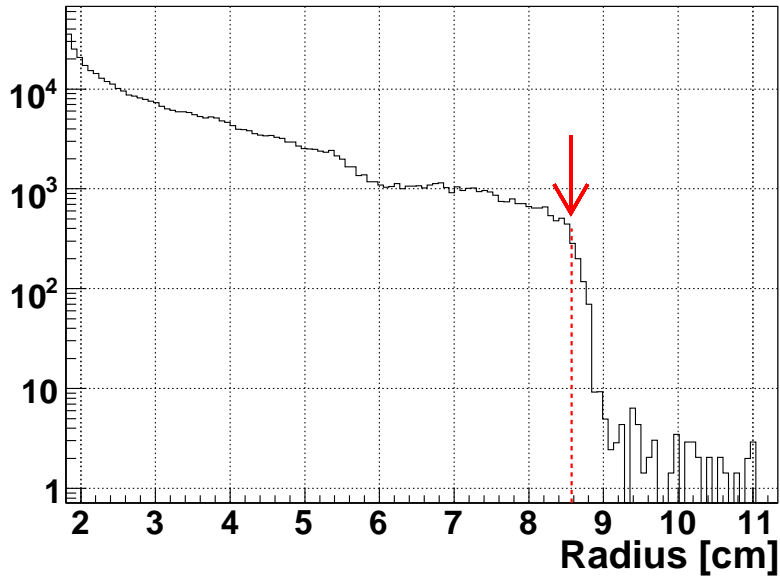}
 \caption{Radial distribution on the pair monitor.
 The shoulder radius, $R_{shl}$ is defined as the radius to
 contain 99.8\% of all the hits, which is shown as an arrow.}
 \label{fig:r}
\end{figure}

\begin{figure}[htbp]\centering
 \includegraphics[width=11cm]{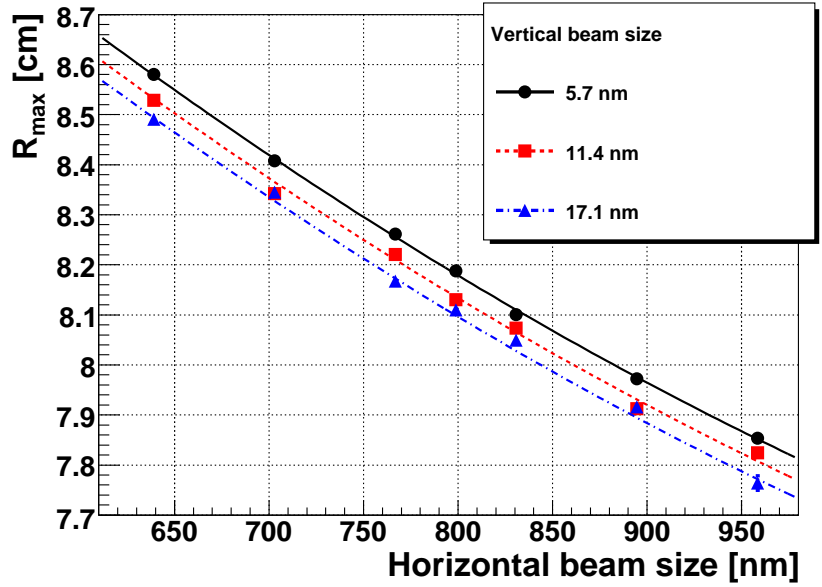}
 \caption{$R_{shl}$ vs. $\sigma_x$ fitted with second order polynomials.
 $R_{shl}$ decreases for larger $\sigma_x$ independent of $\sigma_y$.}
 \label{fig:rmaxVsSx}
\end{figure}

The maximum radius of hit reflects the maximum 
transverse momentum of the pairs, which in
turn is given by the electromagnetic fields of the oncoming
beam. Since the vertical beam size is much smaller than the horizontal and
longitudinal beam sizes, the maximum electromagnetic field is 
inversely proportional to the horizontal and longitudinal
beam sizes. Its dependence on the vertical beam size is negligible for
the ILC beam condition \cite{TauchiAndYokoya}.
Figure \ref{fig:r} shows the radial hit distribution for
the nominal beam bunch crossing which shows a shoulder around 8.6 cm
which corresponds to the maximum transverse momentum.
When the horizontal and/or longitudinal beam size is larger than the nominal
beam size, the position of the shoulder is shifted to a smaller radius. 
We defined the shoulder radius ($R_{shl}$) as the radius to contain 99.8\% of all the hits.
$R_{shl}$ for the nominal beam sizes is shown by the
arrow in Figure \ref{fig:r}.
Figure \ref{fig:rmaxVsSx} shows $R_{shl}$
as a function of the horizontal beam size. As expected, 
$R_{shl}$ decreases for larger horizontal beam size, and it is almost independent of the
vertical beam size. In addition,
$R_{shl}$ becomes smaller than that of nominal beam crossing
for larger longitudinal beam size.

\begin{figure}[htbp]\centering
 \includegraphics[width=9cm]{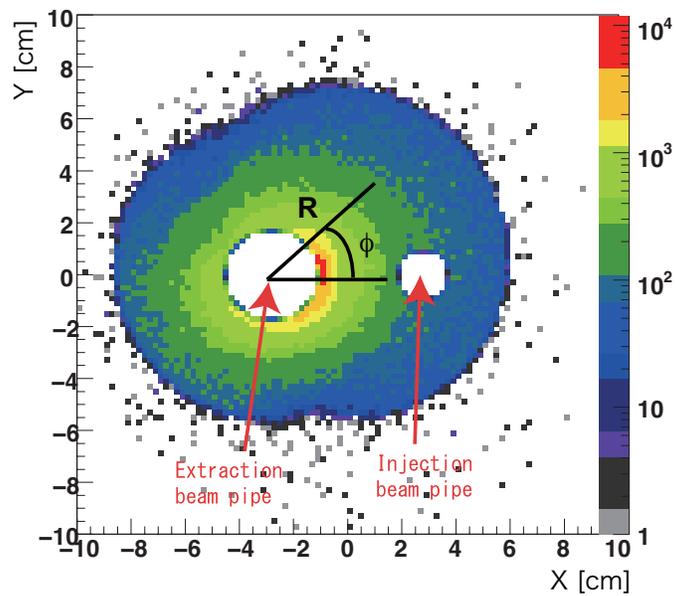}
 \caption{The hit distribution on the pair monitor. There are two holes
 for the incoming and outgoing beams. The radius(R) and the
 azimuthal angle are defined as shown in this figure.}
 \label{fig:xy}
\end{figure}

\begin{figure}[htbp]\centering
 \includegraphics[width=9cm]{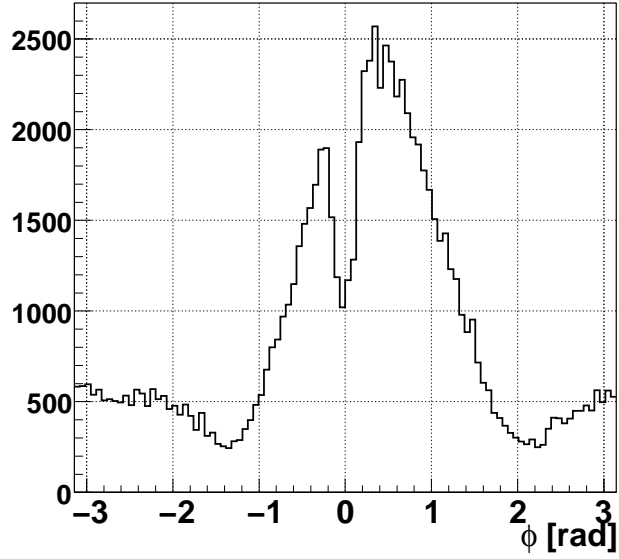}
 \caption{The Azimuthal hit distribution for $R>0.5\cdot R_{shl}$.
The valley at 0 radian is caused by a hole for the incoming beam.
 }
 \label{fig:proj}
\end{figure}

\begin{figure}[htbp]\centering
 \includegraphics[width=11cm]{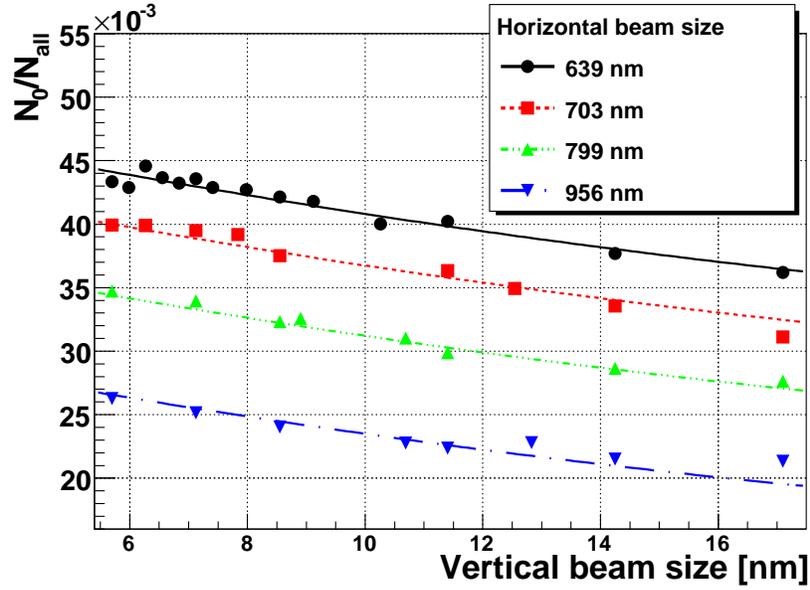}
 \caption{$N_0/N_{all}$ vs. $\sigma_y$ fitted with second order polynomials,
 where $N_0$ is the number of hits in the region defined by $-\pi<\phi<-1.2$ radian and
$2.7<\phi<\pi$ radian for $0.5R_{shl}<R$. $N_0/N_{all}$ decreases for larger
 vertical beam size.}
 \label{fig:ratioVsSy}
\end{figure}

The azimuthal scattering angle of the pairs at the bunch crossing
would depend on the horizontal to vertical aspect ratio of the bunch,
which would then affect the 
azimuthal distribution of the hit density on the pair monitor.
We thus studied
the distribution of the hit density as a function of the radius from the
center of the extraction beam pipe ($R$) and
the angle around the extraction beam pipe ($\phi$).
Figure \ref{fig:xy} shows the hit distribution on the pair monitor, and
Figure \ref{fig:proj} shows the azimuthal hit distribution for $R>0.5\cdot
R_{shl}$. A valley at $\phi=$ 0 radian is due to a hole
on the pair monitor for the incoming beam around $R\sim5.6$ cm and
$\phi\sim 0$ radian. The shape of the azimuthal hit distribution
depends on the radius of the hit distribution around the
extraction beam pipe.
For example, the radius of the right side in Figure
\ref{fig:xy} is larger than that of the left side.
For that reason, we have more events at $\phi\sim0$ in Figure
\ref{fig:proj}. 
We define $N_0$ as the number of hits in $-\pi<\phi<-1.2$ radian and
$2.7<\phi<\pi$ radian for $R>0.5\cdot R_{shl}$. 
In order to derive the beam information from the azimuthal distribution,
we compared $N_0$ to the
total number of hits ($N_{all}$). Figure \ref{fig:ratioVsSy} shows
$N_0/N_{all}$ as a function of the vertical beam size for different
horizontal beam sizes. From this result, $N_0/N_{all}$ is seen to have
information on the
horizontal and vertical beam sizes. This ratio is found to be 
mostly independent of the longitudinal beam size.

In addition, we also use the number of hits $N_1$ in $-1.2<\phi<1.8$ radian for
$R>0.5\cdot R_{shl}$ to increase the sensitivity to the longitudinal beam size. 
The ratio $N_1/N_{all}$ increases for a larger longitudinal beam size, while it decreases
for larger horizontal and vertical beam sizes.

\begin{figure}[htbp]\centering
 \includegraphics[width=11cm]{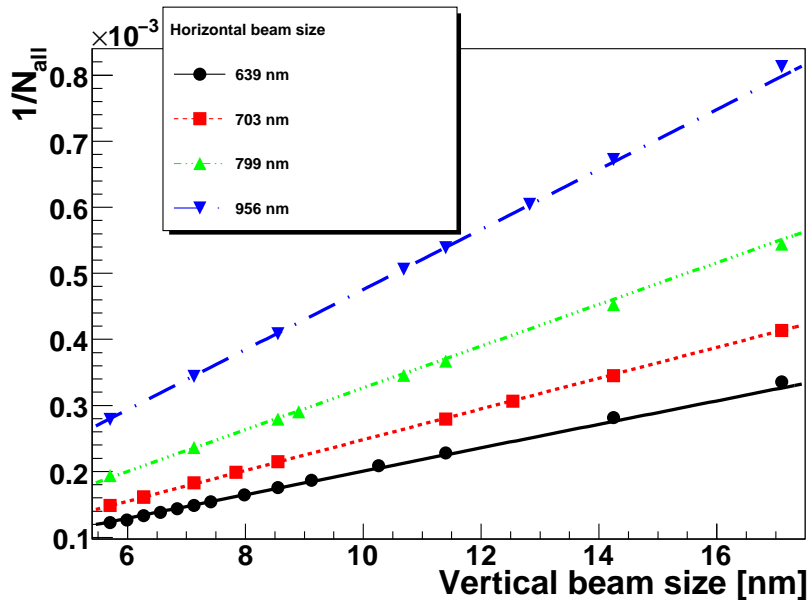}
 \caption{$1/N_{all}$ vs. $\sigma_y$ fitted with second order polynomials. $1/N_{all}$ is inversely
 proportional to the vertical beam size.}
 \label{entries}
\end{figure}

The total number of the hits on the pair monitor, $N_{all}$, reflects the
luminosity which is inversely proportional to the vertical and horizontal beam
size as shown in Equation (\ref{luminosity}) \cite{phd}.
Since the total number of the pair backgrounds are nearly
proportional to luminosity, the number of all the hits on the pair monitor
is expected to be inversely proportional to the vertical and horizontal
beam sizes.
Figure \ref{entries} shows $1/N_{all}$ as a function of the vertical
beam size for several horizontal beam sizes.

\section{Reconstruction of beam sizes}
\label{reconstruction of beam size}
To reconstruct the beam sizes, four measurement variables ($R_{shl}$,
$N_0/N_{all}$, $N_1/N_{all}$, $1/N_{all}$) were used in this analysis. 
Table \ref{tbl:fittingResults} shows the result of fitting each
measurement variable($m_i$) with second order polynomials given by
\begin{eqnarray}
 m_i &=& m_i(\sigma_x^0, \sigma_y^0, \sigma_z^0)
  +\sum_{\alpha=x,y,z} \frac{\partial m_i}{\partial \sigma_\alpha}
  \cdot\sigma_\alpha^0\cdot\frac{\sigma_\alpha-\sigma_\alpha^0}{\sigma_\alpha^0}
  \nonumber \\
 &+&\frac{1}{2}
\sum_{\alpha=x,y,z}\sum_{\beta=x,y,z}
   \frac{\partial^2m_i}{\partial\sigma_\alpha\partial\sigma_\beta}\cdot\sigma_\alpha^0\sigma_\beta^0\cdot\frac{\sigma_\alpha-\sigma_\alpha^0}{\sigma_\alpha^0}\cdot
\frac{\sigma_\beta-\sigma_\beta^0}{\sigma_\beta^0},
\label{fittingFunction}
\end{eqnarray}
where $m_i=R_{shl},\ N_0/N_{all},\ N_1/N_{all},\ 1/N_{all}$.
Each measurement variable in the table is normalized by its value for
the nominal beam sizes; namely, 8.58, 4.43$\times10^{-2}$, 1.62$\times10^{-1}$, and 1.24$\times10^{-4}$
 for $R_{shl}$, $N_0/N_{all}$, $N_1/N_{all}$, and $N_{all}^{-1}$, respectively.
We obtain the numerical values of the matrix ($A_1$) and
the tensor ($A_2$) of Equation (\ref{inverseMatrix}) by fitting the data by 
second order polynomials. 
Then, they were
substituted for Equation (\ref{matrix}) as follows:
\begin{eqnarray}
\left(
\begin{array}{l}
 w_1 \cdot \Delta R_{shl}\\
 w_2 \cdot \Delta N_0/N_{all}\\
 w_3 \cdot \Delta N_1/N_{all}\\
 w_4 \cdot \Delta N_{all}^{-1}
\end{array}
\right)
&=&
\left(
\begin{array}{ccc}
  w_1 \cdot \frac{\partial R_{shl}}{\partial \sigma_x} &
  w_1 \cdot \frac{\partial R_{shl}}{\partial \sigma_y} &
  w_1 \cdot \frac{\partial R_{shl}}{\partial \sigma_z} \\
  w_2 \cdot \frac{\partial N_0/N_{all}}{\partial \sigma_x} &
  w_2 \cdot \frac{\partial N_0/N_{all}}{\partial \sigma_y} &
  w_2 \cdot \frac{\partial N_0/N_{all}}{\partial \sigma_z} \\
  w_3 \cdot \frac{\partial N_1/N_{all}}{\partial \sigma_x} &
  w_3 \cdot \frac{\partial N_1/N_{all}}{\partial \sigma_y} &
  w_3 \cdot \frac{\partial N_1/N_{all}}{\partial \sigma_z} \\
  w_4 \cdot \frac{\partial N^{-1}_{all}}{\partial \sigma_x} &
  w_4 \cdot \frac{\partial N^{-1}_{all}}{\partial \sigma_y} &
  w_4 \cdot \frac{\partial N^{-1}_{all}}{\partial \sigma_z} \\
\end{array}
\right)
\cdot
\left(
\begin{array}{c}
 \Delta \sigma_x\\
 \Delta \sigma_y \\
 \Delta \sigma_z
\end{array}
\right)
\nonumber \\ 
  &+&
   % 2-nd order
 \left(
\begin{array}{c}
 \Delta \sigma_x\\
 \Delta \sigma_y\\
 \Delta \sigma_z
\end{array}
 \right)^T
 \cdot
  \biggl( \mathrm{O}(2) \biggr)
  \cdot
\left(
\begin{array}{c}
 \Delta \sigma_x\\
 \Delta \sigma_y\\
 \Delta \sigma_z
\end{array}
\right).
\end{eqnarray}
The normalization of each measurement variable ($w_1, w_2, w_3, w_4$)
 was adjusted to make the
measurement errors of all the variables numerically equal, namely,
\begin{eqnarray}
 w_1\cdot\Delta R_{shl} = w_2\cdot\Delta N_0/N_{all} = w_3\cdot\Delta
 N_1/N_{all} = w_4\cdot\Delta N_{all}^{-1}.
\end{eqnarray}
This adjustive method is the same as the method of least squares if
there is no correlation between each measurement variable.
The beam size at IP is then reconstructed by the inverse
matrix method. Since we used second order polynomials for the
fitting, we considered up to the second order in Equation
(\ref{inverseMatrix}): 
\begin{eqnarray}
 \Delta\vec{\sigma} = (A_1+\Delta\vec{\sigma}^T\cdot A_2)^+\cdot\Delta\vec{m}.
\end{eqnarray}
This equation is solved iteratively as follows \cite{fastBeamDiagnostics}:
\begin{description}
 \item[(0)] $\Delta\vec{\sigma}_0 = A_1^+ \cdot \Delta\vec{m}$
 \item[(1)] $\Delta\vec{\sigma}_1 = \left[A_1 +
	    \Delta\vec{\sigma}_0^TA_2\right]^+\cdot \Delta\vec{m}$\\
       $\vdots$
 \item[(n)] $\Delta\vec{\sigma}_n = \left[A_1 +
	    \Delta\vec{\sigma}_{n-1}^TA_2\right]^+ \cdot \Delta\vec{m}$\\ 
\end{description}
The iteration was repeated until consecutive iterations satisfied
$$\left(\Delta\vec{\sigma}_n-\Delta\vec{\sigma}_{n-1}\right)
/\Delta\vec{\sigma}_n<1\%.$$
Usually, the number of iteration was 3 to 15.

\begin{table}[htbp]\centering
 \begin{tabular}{|c||c|c|c|c|}
  \hline
  & $R_{shl}$ & $N_0/N_{all}$&
  $N_1/N_{all}$ & $N_{all}^{-1}$ \\ \hline   \hline
%  $m_i(\sigma_x^0, \sigma_y^0, \sigma_z^0)$ & 8.58 & 4.43$\times10^{-2}$ & 1.62$\times10^{-1}$ &
%		  1.24$\times10^{-4}$  \\ \hline 
  $m_i(\sigma_x^0, \sigma_y^0, \sigma_z^0) $ & 1& 1& 1& 1\\ \hline 
  $\sigma_x^0 \cdot \partial m_i/\partial \sigma_x $          &
      -0.20 &
	  -1.0 &
	      -0.078 &
		  1.8 \\ \hline 
  $\sigma_y^0 \cdot \partial m_i/\partial \sigma_y $          &
      -0.0057 &
	  -0.11 &
	      -0.0019 &
		  0.82 \\ \hline 
  $\sigma_z^0 \cdot \partial m_i/\partial \sigma_z $          &
      -0.27 &
	  -0.0075 &
	      0.47&
		  0.54 \\ \hline 
  $\sigma_x^{0\ 2} \cdot \partial^2 m_i/\partial \sigma_x^2 $      &
      0.13 &
	  0.78 &
	      0.083 &
		 2.9  \\ \hline 
  $\sigma_y^{0\ 2} \cdot \partial^2 m_i/\partial \sigma_y^2 $      &
      0.00085 &
	  0.022 &
	      -0.012 &
		 -0.0011 \\ \hline 
  $\sigma_z^{0\ 2} \cdot \partial^2 m_i/\partial \sigma_z^2 $      &
      0.21 &
	  -0.48 &
	      -0.31 &
		  0.26 \\ \hline 
  $2\sigma_x^0\sigma_y^0 \cdot \partial^2 m_i/\partial \sigma_x \partial\sigma_y $&
      0.00015 &
	  0.0090 &
	      -0.011 &
		  1.3 \\ \hline 
  $2\sigma_y^0\sigma_z^0 \cdot \partial^2 m_i/\partial \sigma_y \partial\sigma_z $&
      0.0017 &
	  0.026 &
	      0.0072 &
		  0.35 \\ \hline 
  $2\sigma_z^0\sigma_x^0 \cdot \partial^2 m_i/\partial \sigma_z \partial\sigma_x $&
      0.039  &
	  -0.014 &
	      0.019 &
		  1.4\\ \hline
 \end{tabular}
 \caption{Result of fitting each measurement variable with second order
 polynomials as given by Equation (\ref{fittingFunction}).}
\label{tbl:fittingResults}
\end{table}

\begin{figure}[htbp]\centering
 \includegraphics[width=11cm]{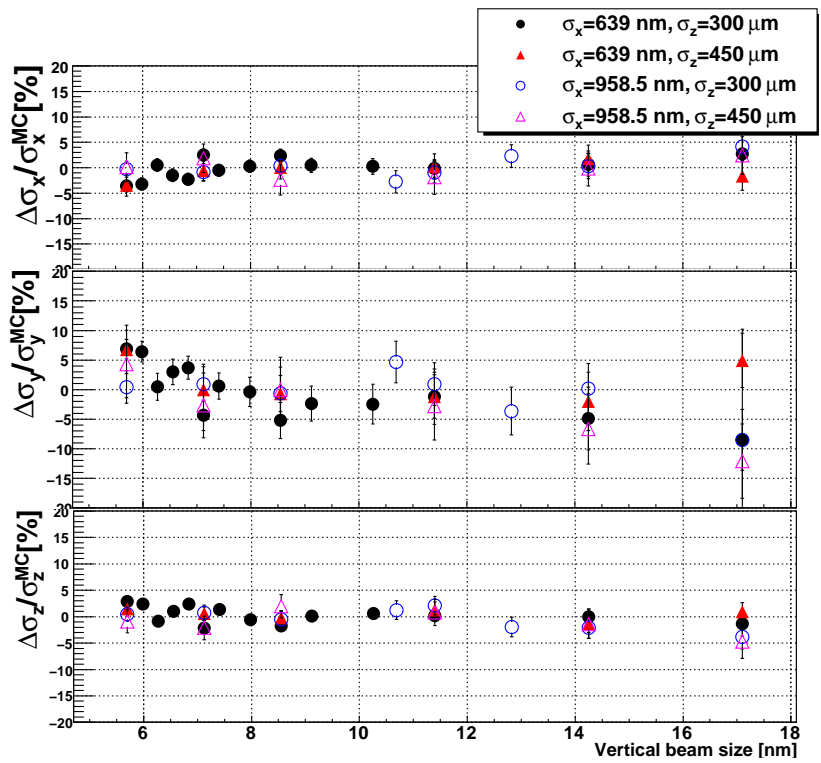}
 \caption{Relative deviations of, from the top, horizontal, vertical and
 longitudinal beam sizes.}
 \label{fig:reso}
\end{figure}

Figure \ref{fig:reso}
shows the relative deviations of the horizontal, vertical, and
longitudinal beam sizes for 50 bunch crossings.
The errors for the distribution of these deviations are estimated at
5.1\%, 10.0\%, and 4.0\% for the horizontal, vertical,
and longitudinal beam sizes, respectively.
Then, we conclude that the pair monitor can measure the beam
sizes with accuracies
of 5.1\% (33 nm), 10.0\% (0.57 nm), and 4.0\% (12 $\mu$m)
for the
horizontal, vertical, and longitudinal beam sizes, respectively.
%It is seen that the pair monitor
%can measure with 5\% (32 nm), 10\% (0.57 nm), and 5\% (15 $\mu$m) accuracy for the horizontal, vertical, and longitudinal beam size, respectively.

\section{Conclusions}
\label{conclusions}
We studied a technique of beam size measurement with the pair monitor.
The method utilizes the second order inversion of the Taylor expansion.
Four measurement variables
($R_{shl}$, $N_0/N_{all}$, $N_1/N_{all}$ and $1/N_{all}$) were used to
reconstruct the beam sizes, and the matrix elements of the expansion
 were obtained
by fitting with second order polynomials of the beam sizes.
The measurement accuracies of the horizontal, vertical, and longitudinal
beam sizes were found to be 5.1\%, 10.0\%, and 4.0\%,
respectively, for 50 bunch crossings.
This result confirms that the pair monitor has, at least statistically, enough
sensitivity to measure the beam size at IP for ILC.

\section*{Acknowledgments}
The authors would like to thank K.Fujii and other members of the
JLC-Software group for useful discussions and helps,
and all the memeber of the FCAL collaboration\cite{FCAL} for all them help.
This work is supported in part by the Creative Scientific Research Grant
(No. 18GS0202) of the Japan Society for Promotion of Science and the
JSPS Core University Program.

% \appendix

% \section{}
% \label{}


\begin{thebibliography}{00}


 \bibitem{RDR}
	 ILC Global Design Effort and World Wide Study, International
	 linear collider Reference Design Report (2007).
 \bibitem{phd}
	 D. Schulte, Study of Electromagnetic and Hadronic Background in the
	 Interaction Region of the TESLA Collider (1996).
 \bibitem{TauchiAndYokoya}
	 T. Tauchi and K. Yokoya, Nanometer Beam-Size Measurement during
	 Collisions at Linear Colliders, KEK preprint 94-122.
 \bibitem{pairCreation}
	 T. Tauchi, K. Yokoya and P. Chen, Pair creation from beam-beam
	 interaction in linear colliders, Particle Accelerator {\bf 41}, 29
	 (1993).
 \bibitem{takubo}
	 Y. Takubo, Proceedings of LCWS 2007,\\
	 \indent\verb+http://www-zeuthen.desy.de/ILC/lcws07/pdf/MDI/takubo_yosuke.pdf+
 \bibitem{Jupiter}
	 Jupiter web-page:\\
	 \indent\verb+http://acfahep.kek.jp/subg/sim/simtools/+
 \bibitem{CAIN}
	 CAIN web-page:\\
	 \indent\verb+http://lcdev.kek.jp/~yokoya/CAIN/cain235/+
 \bibitem{diagnostics}
	 A. Stahl, Diagnostics of Colliding Bunches from Pair Production
	 and Beam-Strahlung at the IP, LC-DET-2005-003.
 \bibitem{fastBeamDiagnostics}
	 M. Ternick, Fast Beam Diagnostics through Beamstrahlung at
	 TESLA.
 \bibitem{FCAL}
	 FCAL collaboration web-page:
	 \indent\verb+http://www-zeuthen.desy.de/ILC/fcal/+

\end{thebibliography}
\end{document}